\RequirePackage{fix-cm}
\documentclass[smallextended]{svjour3}       
\smartqed  
\usepackage{graphicx}
\begin{document}
	
	\title{CLAS12 RICH: New Hybrid Geometry for Strangeness Studies 
	}
	
	
	\author{Giovanni Angelini       \\
		\small(on Behalf of the CLAS and CLAS12 RICH collaborations) 
	}
	
	
	\institute{Giovanni Angelini \at
		Physics Department, The George Washington University, Washington, DC 20052, USA \\
		Tel.: +202-5104440\\    
		\email{gangel@gwu.edu}           
		\and
	}
	
	\date{Received: date / Accepted: date}

	\maketitle
	
	\begin{abstract}

		The Jefferson Lab (JLab) CLAS12 detector will provide a world-leading facility for the study of electron-nucleon scattering. 
		The CLAS12 physics program is very broad and includes studies on quarks dynamics as well as studies on the meson and baryon spectroscopy with quasi-real photoproduction in a large variety of final states.
		The particle identification will be complemented by a Ring Imaging Cherenkov Detector (RICH), that will provide separation of kaons from protons and pions in the momentum range between 3-8 GeV/c allowing the study of pion and kaon electroproduction in semi-inclusive deep inelastic scattering as well as studying  double and triple strange baryons.
		In this article, we will focus on the new technologies developed for the RICH, especially on the complex optical system, that allows to reduce the photo-detection area. Particular attention is given to the characterization of the surface in relationship with the simulation capability of GEANT4.
		
		\newpage
		
	\end{abstract}
	
	\section{Introduction}
	
	\label{intro}
	The CLAS12 spectrometer at JLab will be an unique place for studying electron scattering on nucleon. The high beam intensity, the high beam polarization, the possibility of using longitudinal and transversely polarized targets, and the large acceptance of the detector, will provide data with high statistic and precision. With CLAS12, it will be possible to study the three-dimensional structure of the nulceon via exclusive and semi-inclusive processes. In addition, thanks to the possibility of obtaining quasi-real photons, it will be possible to study the $\Omega$ and the  $\Xi$ photoproduction, and to search for exotic mesons.\\ 
	The CLAS12 baseline particle identification comprises a time-of-flight system and two threshold Cherenkov counters, installed in six azimuthal sectors. However, in its standard configuration CLAS12 did not provide the necessary kaon particle identification capabilities in the full momentum range explored. \\
	In the deep inelastic kinematic region, the yield of pions is expected to be one order of magnitude larger than kaons. For this reason, a pion rejection factor of about 1:500 is required to limit the pion contamination in the kaon sample to a few percent. This is going to be achieved by replacing two of the six sectors of a threshold Cherenkov detector\footnote{The Low Threshold Cherenkov Counter (LTCC)} with a RICH detector that will supply the particle identification in the momentum range between 3 and 8 GeV/c.  The installation of the RICH detector will not only improve the study of the 3D structure of the nucleon via inelastic scattering, but will have also an impact on the studies of the $\Xi$ photoproduction since about 40\% of the kaons obtained from the $\Xi$ reaction will have a momentum larger than 3 GeV/c. These data  will greatly improve the knowledge of the cross section for photoproduction of $\Xi$ baryon, enable measurement of the polarization of the produced $\Xi$ and provide a determination the mass splittings of ground and excited cascades.
	
	\section{The CLAS12 RICH detector}
	\label{sec:1}
\begin{figure}[h!]
		\includegraphics[width=1\textwidth]{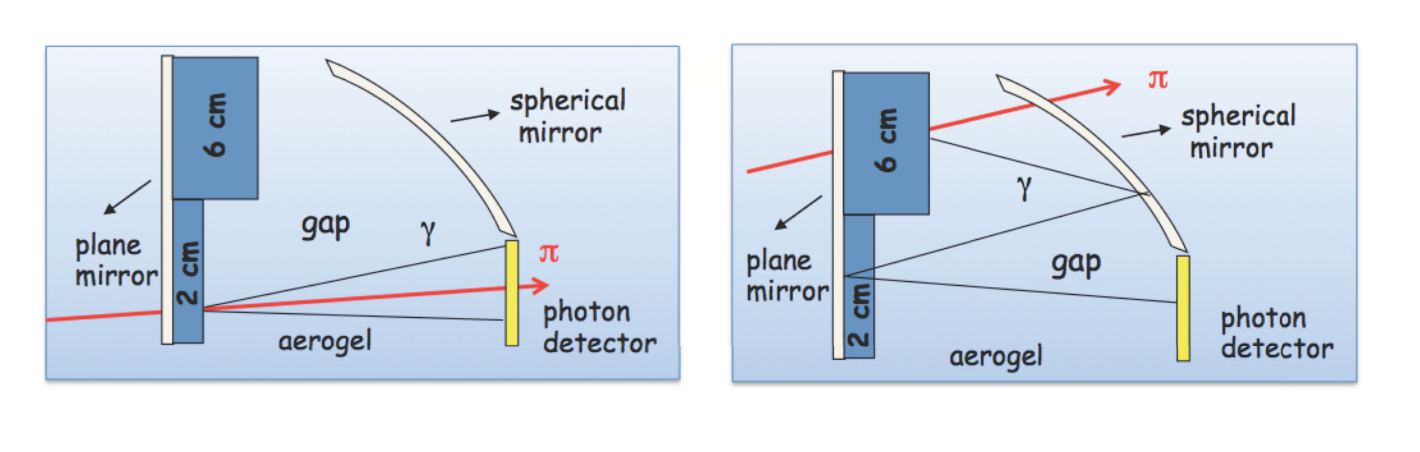}
		\caption{The CLAS12 RICH geometry concept (see text for details).}
		\label{geometry}       
	\end{figure}

	The CLAS12 RICH uses a novel hybrid geometry \cite{Montgomery:2013hva} to overcome the constraints imposed by the existing CLAS12 detector. This non conventional design (Fig.\ref{geometry}) allows to reduce  the area instrumented with photodetectors. For particles with an angle $\theta \leq$  $13 ^\circ$ with respect to the beam direction, the Cherenkov radiation emitted in a layer of aerogel of thickness 2 cm, is directly measured by a series of PhotoMultiplier Tubes (PMTs) positionated at a distance of about 1 m. For larger incident angles of 13$^\circ < \theta < 25 ^\circ$, the Cherenkov radiation is emitted in two layers of aerogel, each having 3 cm of thickness, and the radiation is reflected and converged to the detection area by a complex system of planar and spherical mirrors.\\
	In the momentum range of interest (3-8 GeV/c), the aerogel is the best radiator available. It is an amorphous solid network of SiO$_2$ nanocrystals with a low macroscopic density. The emission spectrum is mostly in the UV and visible range and for that reason PMTs have been the choice for photons detection. In addition, the photon detector must provide a spatial resolution of about 5 mm not to degrade the Cherenkov angle resolution and has to work in the Single Photon-Electron (SPE) regime. These requirements imposed the choice of Hamamatsu MultiAnode PhotoMultipliers Tubes (MAPMTs) H8500 and H12700 \cite{Hoek:2014gha,Mirazita:2017vav}. \\
	Test beams studies with a larger prototype and a simplified mirror system have been performed at the T9 experiment hall at CERN, in which a clear $\pi/K$ separation has been observed in the momentum range of interest \cite{Baltzell:2015fea}.  
	
	\section{The RICH Optical System}
	\label{sec:2}
\begin{figure}[h]
		\includegraphics[width=0.85\textwidth]{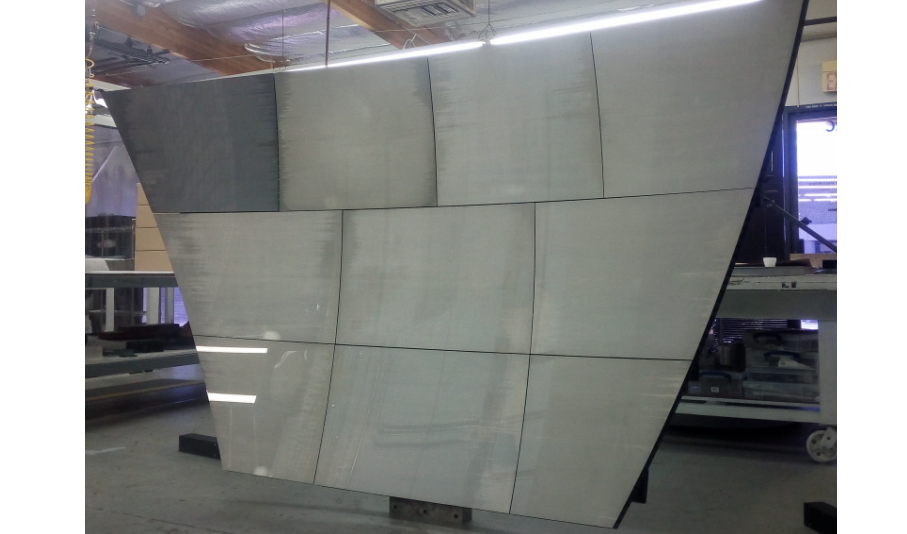}
		\caption{The 10 spherical mirrors before being coated. The spherical mirrors are realized with a carbon fiber skin and a 25 mm core of honeycomb.}
		\label{spherical}       
	\end{figure}

	The optical system of the RICH  is made of seventeen mirrors, ten of them are spherical while the rest are planar. The spherical mirrors (Fig. \ref{spherical}) are realized with carbon fiber layer and a core of honeycomb in order to reduce the material budget of the detector. The planar mirrors are composed of glass in order to reduce the cost while keeping good performances.\\
	Simulation studies and the prototype tests have shown that the required PID capabilities can be reached with a Cherenkov angle resolution of the order of 4 mrad.  In order to maintain the particle identification expected the optical system should introduce a systematic of the order of 1 mrad for the Cherenkov angle reconstruction. Several stand-alone simulations have been realized to characterize the effect of the mirror's gravity load and of the surface topology  otherwise not implemented in GEANT4 \cite{GEANT4} (the simulation software used for describing the detector).  Tests have been performed on prototypes in order to select the best surface topology. In the next subsections more information are given on the simulation strategy and the measurements performed.
	
	\subsection{Simulating the surface contributions }
\begin{figure}
		\includegraphics[width=1\textwidth]{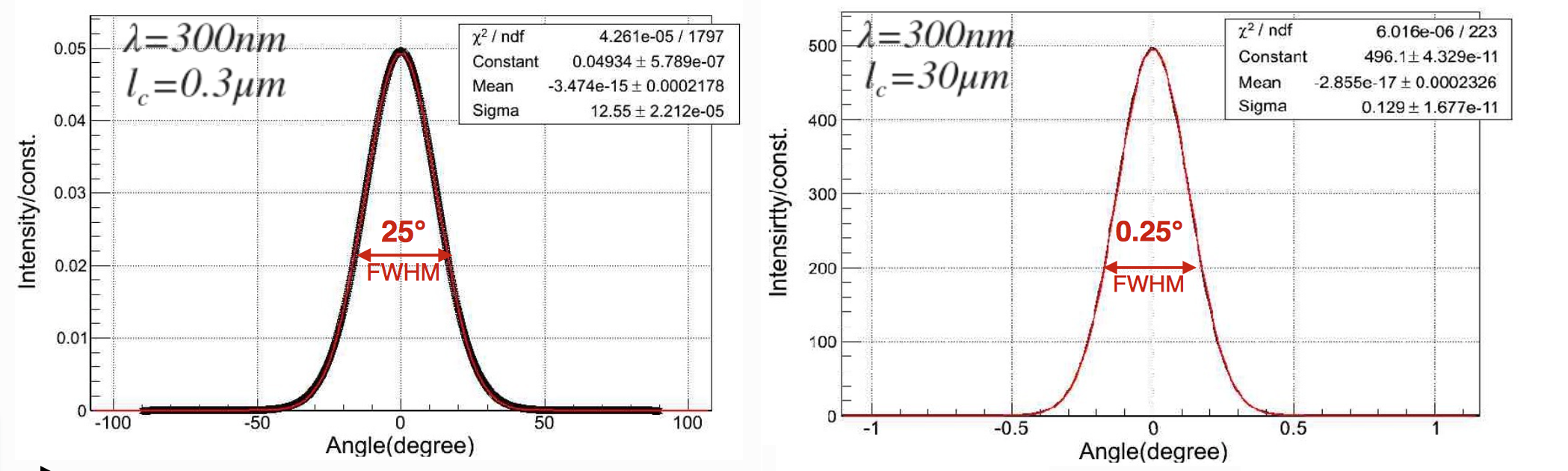}
		\caption{ The photons angular distribution obtained from the Rice model. The intensity is normalized to an arbitrary value. Left: simulations performed with micro-roughness $\sigma_r=3\, nm$, incident light wavelength $\lambda=300\,nm$, and  correlation length $l_c = 0.3 \,\mu m$. The FWHM of the distribution is 25 degrees. Right: simulations performed with micro-roguhness $\sigma_r=3 nm$, incident light wavelength $\lambda=300\,nm$, and correlation length $l_c = 30 \, \mu m$.  The FWHM of the distribution degress to 0.25 degrees. }
		\label{simulation}       
	\end{figure}

\begin{figure}
		\includegraphics[width=0.9\textwidth]{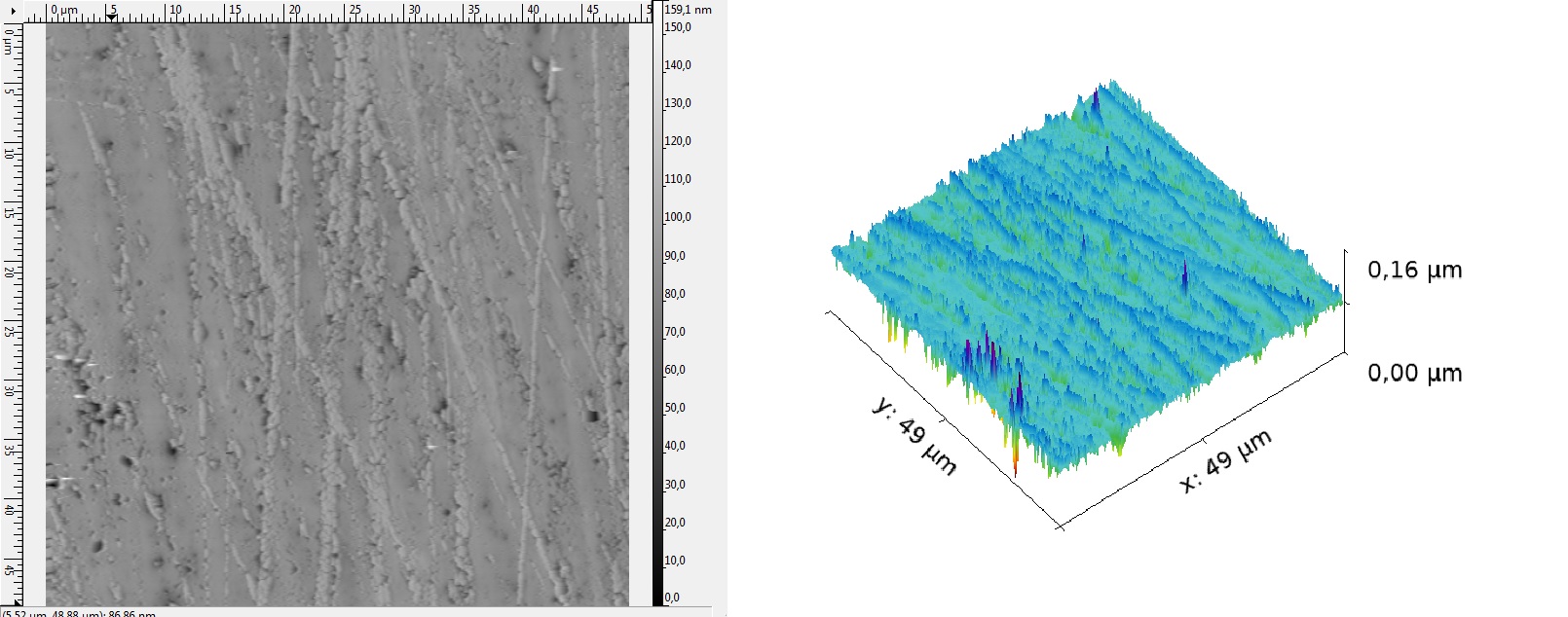}
		\caption{The surface of the carbon fiber spherical mirror performed with an Atomic Force microscope. The data collected were used to compute the PSD (see text for details).}
		\label{AFM}       
	\end{figure}

\begin{figure}
		\includegraphics[width=1\textwidth]{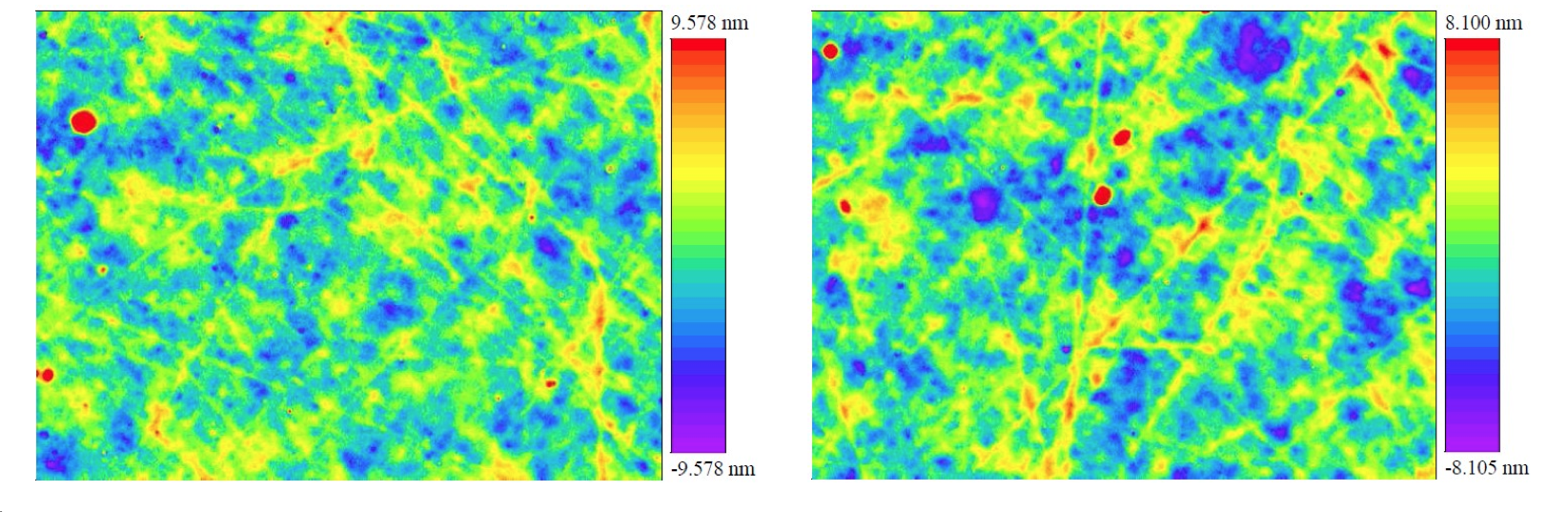}
		\caption{Two measurements performed on the surface of a mirror's prototype by using a white light interferometer. The dimension of x and y are respectively $150 \, \mu m$ an $110 \, \mu m$.  }
		\label{WLI}      
	\end{figure}

	Simulations of complex detectors like the CLAS12 RICH are performed by using the GEANT4 package. Here, the scattering of photons on a rough surface is treated in a  unified model \cite{knoll}. The model uses the so-called Kirchoff approximation, and it requires the only knowledge of the root mean square of the height's fluctuation of the surface with respect to its expected shape, often called roughness ($\sigma_r$).  However, the Kirchoff approximation can be justified only if the roughness is bigger than the wavelength  of the incident radiation. When the roughness is smaller than the wavelength, the Rice model \cite{harvey2007unified,Bechkmann} can be used. This model is valid if \cite{Rice}:
	\begin{equation}
	\left(\frac{4 \pi \sigma_r \, cos(\theta_i)}{\lambda} \right)^2 < \, 1 ,
	\label{microroughness}
	\end{equation}
	where $\lambda$ is the wavelength of the incident radiation, and $\theta_i$ is the angle of incidence of the photon with respect to the normal of the surface. In this article, I will refer to micro-roughness in case $\sigma_r$ satisfies the previous inequality. \\
	Under the following condition \cite{Petit}:
	\begin{equation}
	\frac{\sigma_r}{l_c} \le 2.3 \times 10^{-2}\, ,
	\label{limit}
	\end{equation}
	the model gives simplified equations for the distribution of reflected and scattered radiation. In Eq.(\ref{limit}) $l_c$ is the correlation length of the height's fluctuations along the surface of the mirror. 
	The radiation, i.e., the amount of photons in the specular angle, is given by :
	\begin{equation}
	R_s (\theta_i) = R_0 \, exp[-\left(\frac{4 \pi \sigma_r cos(\theta_i)^2}{\lambda}\right)^2] ,
	\label{specular}
	\end{equation}
	where $R_0$ represents the expected reflectivity under the Fresnel's approximation and depends on the dialectics used  for the coating of the mirrors. The radiation scattered out of the specular angle, i.e. diffused, is given by:
	\begin{eqnarray}
	R'(\theta_s,\phi_s,\theta_i,\phi_i)= \frac{16 \pi^2}{\lambda^4}cos(\theta_i) cos^2(\theta_s) Q(\theta_i) \sigma_r^2 l_c^2\times PSD(\theta_s,\phi_s,\theta_i,\phi_i)
	\label{reflected}
	\end{eqnarray}
	where, $\theta_i$, $\phi_i$, $\theta_s$, $\phi_s$ represent the incident and reflected polar and azimuthal angles, Q($\theta_i$) is the angle dependence polarization, and PSD is the Power Spectral Density function of the surface \cite{Noll}.\\
	We generated $10^4$ surfaces with different micro-roughness and correlation length (Gaussian distributed) and we simulated the effect on the Cherenkov angle resolution by implementing the Rice model in a stand-alone code (Fig. \ref{simulation}).  We identified that for surfaces having $\sigma_r \le 4 \, nm$ and $l_c  > 50 \,  \mu m$ the contribution of the Cherenkov angle was $ < $(0.5 $\pm$ 0.1) mrad\footnote{The error has been obtained as variance between $10^4$ simulations obtained at fixed $\sigma_r$ and $l_c$}. \\
	While these constraints were easy to obtain on the planar glass mirrors, it was challenging to obtain a carbon fiber surface within these limits and several measurements were performed on prototypes realized by different firms. For this reason, an Atomic Force Microscope (AFM), capable of resolving fluctuations of the order of $nm$, and  a White Light Interferometer (WLI), able to measure fluctuations  from $10 \, nm$ to $10 \, \mu m$, were used to measure the micro-roughness of the mirror's prototypes. As an example, Fig. \ref{AFM} and Fig. \ref{WLI} show two measurements realized with the AFM and WLI. These measurements allowed to select the mirrors produced by the Composite Mirror Application Inc.\footnote{http://www.compositemirrros.com/} as the best for the CLAS12 RICH. \\
	Once the surface topology was defined, we computed the effect of the roughness at macroscopic scale by realizing simulations with a ray-tracing algorithm. The results showed that a roughness smaller than 1 $\mu$m was necessary to maintain the contribution to the Cherenkov resolution below 1 mrad.\\
	The validation of these studies was obtained by measuring the reflected spot when the mirror is illuminated by a point-like source from its Center Of Curvature (COC). In case of a perfect mirror, the intensity distribution is described by a delta distribution, and consequently the measurement of the spot size can be related to the surface imperfections. The spot size is defined by the diameter containing $95\%$ of the total intensity, the so-called D$_0$ of the spot. We measured the first prototypes and obtained a  D$_0$=4.6 mm, in agreement with the simulations result of (5$\pm$0.5) mm.

	\subsection{Gravity load studies}
	Another important effect that can contribute  to the optical system performance, is the deformation of the optical mirrors due to the stress of the gravitational force on the mirror structure. Simulations were performed using the commercial software ANSYS\footnote{https://www.ansys.com} . This software allows the identification of the displacements that a mechanical structure undergoes because of the gravity load and thermal deformation, using a Finite Element Method (FEM) \cite{ANSYS}. At the result of the FEM we applied a ray tracing algorithm to define the effect on the direction of the reflected Cherenkov photons.\\
	The deformations were  negligible (with only 0.1 mrad of contribution), for an honeycomb core of  thickness between 2-3 cm, resulting in a material budget of the order of $0.1$ radiation lengths. Similar result were obtained on the glass mirrors.
	
	\section{Conclusion}
	The CLAS12 RICH will enhance the particle identification capability of the CLAS12 spectrometer in a wide kinematic range. The particle identification quality achievable with the RICH will allow for precise studies of the strangeness component of the nucleon. Simulations with stand-alone codes, together with  experimental measurements, have been performed in order to characterize the complex optical system of the detector. This system has recently been completed, and installed in the detector.
	\paragraph{}
	\paragraph{}
	\textit{This material is based upon work supported by the U.S. Department of Energy, Office of Science, Office of Nuclear Physics under contract DE-AC05-06OR23177.}
	
	%

	%

	
	

\end{document}